\documentclass[12pt,epsf]{article}
\newcommand{\be}{\begin{equation}}
\newcommand{\ee}{\end{equation}}
\newcommand{\bea}{\begin{eqnarray}}
\newcommand{\eea}{\end{eqnarray}}
\newcommand{\eps}{\epsilon}

\newcommand{\Pm}{{\cal P}_-}
\newcommand{\Ppm}{{\cal P}_\pm}

\newcommand{\T}{\Theta}
\newcommand{\ksym}{$\kappa$-symmetry }

\begin{document}
\begin{flushright}
\hfill{SU-ITP-98/47}\\
\hfill{hep-th/9808038}\\
\hfill{August 1998}\\
\end{flushright}

\vspace{20pt}

\begin{center}
{\large {\bf The GS String Action on $AdS_5\times S^5$}}

\vspace{40pt}

{\bf Renata Kallosh$^{a}$ and J. Rahmfeld$^{b}$}

\vspace{20pt}
{\it Department of Physics,

Stanford University,

Stanford, CA 94305-4060}

\vspace{60pt}

\underline{ABSTRACT}

\end{center}

We present a simple form of the Type IIB string action on $AdS_5\times S^5$. The
result is achieved by fixing $\kappa$-symmetry in the Killing spinor gauge
defined by the projector of the Killing spinor of the D3 brane. We show
explicitly that in this gauge the superspace is greatly simplified which is the
crucial ingredient for the simple string action.

{\vfill\leftline{}\vfill
\vskip  30pt
\footnoterule
\noindent
{\footnotesize
$\phantom{a}^a$ e-mail: kallosh@physics.stanford.edu. }  \vskip  -5pt

\noindent
{\footnotesize
$\phantom{b}^b$ e-mail: rahmfeld@leland.stanford.edu. }  \vskip  -5pt

%\vskip  14pt

%\baselineskip=24pt
\pagebreak
\setcounter{page}{1}

Recently, research in worldvolume theories of strings and extended objects on
$AdS_*\times S^*$ backgrounds has attracted attention, motivated by the
conjectured duality between $D=4$ Large $N$ Yang-Mills theory and Type IIB
string theory compactified on $AdS_5 \times S^5$ \cite{Malda, Klebanov, Witten}.
 Clearly, it is of great importance to obtain the string action in
this background in a reasonably simple form, which is the purpose of this note.

In flat superspace, the string action with manifest ten dimensional super
Poincare symmetry was discovered by Green and Schwarz \cite{GS}. The classical
action is non-linear. Upon gauge-fixing in the light-cone gauge the gauge-fixed
theory is free as the action is quadratic. The GS string action was generalized
in \cite{Grisaru} to generic Type IIB backgrounds. In such backgrounds in which
even after gauge-fixing the action is not free.

More recently, the classical string action was presented in a closed form in
$AdS_5
\times S^5$ supercoset construction \cite{Tseytlin, NearHorizon,Tseyt3}:
\begin{eqnarray}
S =-\frac{1}{2}\int d^2\sigma\ \left[\sqrt{g} \, g^{ij}
 L_i^{\hat a} L_j^{\hat a} +  4 i \eps^{ij}\int_{0}^1 ds
 L_{is}^{\hat a} S^{IJ} \bar \Theta^I \Gamma^{\hat a} L_{is}^{J}
 \right]
\ ,
\label{action}
\end{eqnarray}
where $S^{IJ}$ has non-vanishing elements $S^{11}= -S^{22}=1$ and where $\hat
a=0,..,9$.

\begin{equation}
L_s^{ I} =\left[ \left({\sinh \left({s\cal M}\right) \over {\cal M}}\right)
D\Theta
\right]^{I}
\label{LI}
\end{equation}
and
\begin{eqnarray}
L_s^{\hat a }&=&e^{\hat a }_{\hat m} (x) dx^{\hat m} - 4 i \bar
\Theta^I\gamma^{\hat a}
\left({
\sinh^2  \left({s\cal M}/2\right) \over {\cal M}^2} D\Theta \right)^I\label{La}
\end{eqnarray}
where
\begin{eqnarray}
({\cal M}^2)^{ IL}&=& [ \epsilon^ {IJ} (-\gamma^{ a} \Theta^{J} \bar
\Theta^L \gamma^{ a} + \gamma^{ a'}
\Theta^{J} \bar \Theta^L \gamma^{ a'} )\nonumber\\
&+& {1\over 2}
\epsilon^{KL} (\gamma^{ab} \Theta^I \bar \Theta^K \gamma^{ab}
-\gamma^{a'b'} \Theta^I \bar \Theta^K \gamma^{a'b'})
], \label{msquare}
\end{eqnarray}
and $L^I=L^I_{s=1}, \ L^{\hat a}=L^{\hat a}_{s=1}$. Here,
\begin{equation}
(D\Theta)^I =\left ( d +{1\over 4}( \omega^{ab} \gamma_{ab} +
\omega^{a'b'}
\gamma_{a'b'}) \right )\Theta^I  -{1\over 2} i\epsilon^{IJ} (e^a
\gamma_a +
ie^{a'} \gamma_{a'}) \Theta ^J \label{DTheta}
\end{equation}
with $d=dx\partial_x +d\Theta\partial_\Theta$. Also, we use a 5+5 split
\cite{Tseytlin} of $\hat a$ into $\hat a=(a,a')$. This presents a closed form of
a $\kappa$- and reparametrization symmetric string action in $AdS_5\times S^5$
background. Although closed, the classical action as it is depends on even
powers of $\Theta$ up to $\T^{32}$.

Since both symmetries are local, they have to be gauge-fixed. As shown in
\cite{SuperKilling} there exists a gauge which utilizes the Killing spinors of
the background which leads to a significant simplification of the action. The
argument was based on the properties of a supersolvable subalgebra of the full
superconformal algebra. The connection between gauge fixing \ksym and
supersolvable subalgebras was discovered in \cite{Fre} in the context of the
$M2$-brane.

The program of gauge-fixing consists of these steps:
\begin{itemize}
\item The choice of an algebraic projector which eliminates 1/2 of the 32
  spinorial degrees of freedom of the classical action. This projector will
  be suggested by the full $D3$-brane Killing spinor. The surviving fermionic
  directions will be precisely those Killing directions of the geometry.

\item A change of the fermionic coordinates to accommodate the space-time
 dependence of the Killing spinors.

\item Ensuring the consistency of the gauge.

\end{itemize}

The split of the 32 component spinor of the classical string action is suggested
by the Killing spinor of the full $D3$-brane background which preserves $1/2$ of
the supersymmetries. Define the projectors
\be
\Ppm=\frac{1}{2}\left(\delta^{IJ} \pm \Gamma_{0123}\eps^{IJ}\right),
\ee
where $\Gamma_{0123}$ is the product of $\Gamma$ matrices in the direction of
the brane. The gauge is defined by
\be
\Pm \Theta=0. \label{fixer}
\ee
For future convenience we introduce besides the 5+5 split $\hat a=(a,a'), \hat
m=(m,m')$ also the 4+6 split $
\hat a= (\bar p,\bar t), \hat m=( p,t)$ with $p,\bar p=0,1,2,3$
and $t,\bar t= 4,..,9$. This reflects the separation of the space-time
coordinates into directions along the $D3$-brane and transverse to it. With
\be
\Theta_{\pm}= \Ppm \Theta
\ee
we can define a convenient basis
\bea
\Theta^1_\pm \equiv&(\Theta_\pm)^1& \equiv \frac{1}{2}\left(
\Theta^1 \pm \Gamma_{0123}\Theta^2
  \right) \nonumber \\
\Theta^2_\pm \equiv&(\Theta_\pm)^2 &\equiv \frac{1}{2}\left(\Theta^2
\mp \Gamma_{0123}\Theta^1
\right)=\mp \Gamma_{0123} \Theta^1_\pm .
\eea
In this basis the gauge reads
\be
\Theta_-^1=\Theta_-^2=0.
\ee
A tremendous simplification of the supervielbeins occurs if we apply this gauge
to (\ref{LI}) and (\ref{La}). Notice that in these objects all fermionic
contributions appear in the form
\be
{\left({\cal M}^{2n}\right)} D\Theta.
\label{form}
\ee
We will show below that
\be
{\left({\cal M}^{2}\right)} D\Theta_+=0
\ee
which means that all terms of the type (\ref{form}) vanish for $n>0$. For
simplicity, we prove this in the 5+5 split.

First, rewrite $\Theta_-=0$ as
\be
\Theta_+^I=i\eps^{IJ}\gamma^{4}\Theta^J_+
\label{Cons5}
\ee
where we used
\be
\Gamma_{0123}=i\gamma^4\times 1_{2\times 2}.
\ee
In general, the relation between ten dimensional Dirac matrices $\Gamma$ and 5+5
dimensional ones $\gamma^a$ and $\Gamma^{a'}$ are \cite{Tseytlin}
\bea
\Gamma^{a}&=&\gamma^a\times \sigma_1, \nonumber \\
\Gamma^{a'}&=&\gamma^{a'}\times \sigma_2,
\eea
where $\gamma^a$ and $\gamma^{a'}$ obey the (anti-)commutation relations
\be
\left\{\gamma^a,\gamma^b  \right\}=2\eta^{ab}, \qquad
 \left\{\gamma^{a'},\gamma^{b'}  \right\}=2\delta^{a'b'}, \qquad
  \left[\gamma^a,  \gamma^{b'}\right]=0.
\ee
With (\ref{Cons5}) we arrive at
\begin{eqnarray}
({{\cal M}_{\rm fix}}^2)^{ IL}D\Theta_+^L&=& \bigl[
(i\gamma^{a}\gamma^4\Theta_+^{I}\bar \Theta_+^L \gamma^{a} D\Theta_+^L
 -i\gamma^{a'}\gamma^4\Theta_+^{I}\bar \Theta_+^L \gamma^{a'} D\Theta_+^L)
 \nonumber\\
&+& {1\over 2}\bigl(
-i \gamma^{ab} \Theta_+^I \bar \Theta_+^K \gamma^{ab}
  \gamma^4 D\Theta^K
+i \gamma^{a'b'} \Theta_+^I \bar \Theta_+^K \gamma^{a'b'}
  \gamma^4 D\Theta_+^K\bigr)
\bigr]. \label{Mcons}
\end{eqnarray}
To show that this expression vanishes we use the fact that all terms of the
structure
\be
\bar \Theta_+^I \hat \Gamma D\Theta^I \quad {\rm with} \quad
\left[\gamma^4,\hat \Gamma\right]=0
\ee
vanish. This leads to
\be
({{\cal M}_{\rm fix}}^2)^{IL}D\Theta_+^L= \bigl[
(i\gamma^{p}\gamma^4\Theta_+^{I}\bar \Theta_+^L \gamma^{p} D\Theta_+^L) +
{1\over 2}\bigl(
-i \gamma^{4p} \theta^I \bar \Theta_+^K \gamma^{4p}
  \gamma^4 D\Theta_+^K
\bigr)
\bigr]=0, \label{main}
\ee
as can be easily verified. We are left after gauge-fixing with the following
supervielbeins:
\bea
(L_{s}^{ I})_+ &=&s D\Theta_+^I \nonumber \\ (L_{s}^{I})_- &=&0
\nonumber \\
L_s^{p}&=&e^{p}_{\hat m} (x) dx^{\hat m} - i s^2 \bar
\Theta_+^I\gamma^{p} D\Theta_+^I\label{La2} \\
L_s^{t}&=&e^{t}_{\hat m} (x) dx^{\hat m}, \nonumber
\end{eqnarray}
where the indices $p$ and $t$ refer to directions parallel and orthogonal to the
$D3$-brane.

\vspace{1cm}
\noindent
We now turn to the change of fermionic variables. At this point it is useful to
remind ourselves of the explicit $AdS_5\times S^5$ metric. Since in the
following we use both notations, $5+5$ and $6+4$ split, we give it in
spherical/AdS coordinates as well as in Cartesian coordinates:
\bea
{\rm spherical/AdS:}& \qquad &ds^2=r^2(dx^p dx_p)+\frac{dr^2}{r^2} +d \Omega^2
\\
{\rm Cartesian:}& \qquad &ds^2=y^2(dx^p dx_p)+\frac{1}{y^2}(dy^t dy_t).
\eea
To simplify the expression $D\Theta_+$ the following change of variables (in
Cartesian coordinates) is in order\footnote{In spherical coordinates the
suitable change of variables is
\be
\Theta_+^I= f(\eta) g_{tt}^{\frac{1}{4}}(r) \theta_+^I,
\label{spherical}
\ee
where $f(\eta)$ is a function of the 5 Euler angles of the
5-sphere\cite{KillingSpinors}.}:
\be
\Theta_+^I= g_{tt}^{\frac{1}{4}}(|y|) \theta_+^I \label{NewCoord}
\ee
With this definition $D\Theta_+^1$ simply becomes \cite{NearHorizon}
\be
D\Theta_+^1=(g(|y|))^{\frac{1}{4}} d\theta_+=\sqrt{|y|} d\theta_+.
\ee
This can be seen by enforcing the constraint (\ref{fixer}) on (\ref{DTheta})
which reduces to solving the Killing spinor equation for the full D3-brane
Killing spinor in the near horizon regime.

In the coordinates (\ref{NewCoord}) the supervielbeins take the form (at $s=1$)
\bea
L_+^{ I} &=&\sqrt{|y|} d\theta_+^{I} \nonumber \\ L_-^{ I} &=&0 \nonumber \\
L^{p}&=&|y|
 (dx^p -  i \bar \theta_+^I \Gamma^{p} d\theta_+^I)\\
 L^{t}&=&\frac{1}{|y|} dy^t
\eea
This allows us to reduce the complicated classical action to a much simpler
gauge-fixed action. With the obvious replacements $dZ^M \rightarrow\partial_i
Z^M d\sigma_i$ we find
\begin{eqnarray}
S =-\frac{1}{2}\int d^2\sigma\ \biggl[\sqrt{g} \, g^{ij}&& \hspace{-0.7cm}
\left(y^2(\partial_i x^p
- 2 i \bar
\theta_+ \Gamma^{p} \partial_i\theta_+)(\partial_j x_p - 2 i \bar
\theta_+ \Gamma_{p} \partial_j \theta_+) +\frac{1}{y^2} \partial_i y^t
\partial_j y^t \right) \nonumber \\ &&
 + 4 i \eps^{ij} \partial_i y^t (\bar \theta_+ \Gamma^t \partial_j\theta_+)\biggr]
\label{SimpleAction}
\end{eqnarray}
which constitutes our main result\footnote{If we would work instead in spherical
coordinates, we would change the $y^2$ to $r^2$, modify the second term to
$\frac{1}{r^2}
\partial_i r \partial_j r+...$ (where the dots denote derivatives
on angular coordinates), and include the angular dependence of the spinors as in
(\ref{spherical}).}. Here, for compactness $\theta_+$ denotes $\theta_+^1$. The
last term originates from the WZ term.

The action still remains reparametrization invariant. There are various
possibilities to fix this symmetry which have to be studied of the $AdS_5
\times S^5$ background.

\vspace{1cm}

Finally, one has to ensure that the \ksym gauge (\ref{fixer}) is an acceptable
one, which requires the differential operator in the quadratic part of the
fermions to be invertible. The relevant term in the action is of the form
\be
{\cal L}\sim \bar \theta_+ \left[(\Pi_{p} \Gamma^p +\Pi_{t}
\Gamma^t)_z
\partial_{\bar z} + (\Pi_{p} \Gamma^p -\Pi_{t} \Gamma^t)_{\bar z}
\partial_ z\right] \theta_+.
\ee
The zero modes of the classical action in this notation are
\be
(\Pi_{p} \Gamma^p +\Pi_{t}
\Gamma^t)_z \qquad {\rm and } \qquad
(\Pi_{p} \Gamma^p +\Pi_{t}
\Gamma^t)_{\bar z}.
\ee
None of them remains a zero mode in our gauge fixed action, hence the gauge we
have used is acceptable.

In summary, we have shown, following \cite{SuperKilling}, that $\kappa$-symmetry
of the Green-Schwarz string can be used to remove fermionic degrees of freedom
in a way that simplifies the $AdS_5 \times S^5$ superspace geometry and
therefore the action.

\vskip 2 cm
We had stimulating discussion with Arvind Rajaraman and Arkady Tseytlin. The
work of R.K and J.R is supported by the NSF grant PHY-9219345.

\vskip 2 cm
\noindent
{\large \bf Note Added:}

After completion of this work we became aware of the paper by I. Pesando
\cite{Pesando} which displays the string action with a simple fermionic sector.
The action was obtained via the technique of supersolvable algebras, and the
relation to gauge fixing procedure still needs to be fully understood.

%\bibliographystyle{preprint}
%\bibliography{AdS5}

\end{document}